\documentclass[review,12pt,authoryear]{elsarticle}
\usepackage{microtype}
\DisableLigatures[f,s,c,?,!]{encoding = *, family = * }

\usepackage{amssymb}
\usepackage{array}
\usepackage{cancel}
\usepackage{lineno}
\usepackage{mathtools}
\usepackage{multirow}
\usepackage[super]{nth}
\usepackage{physics}
\usepackage{rotating}
\usepackage{setspace}
\usepackage{soul}
\usepackage[inline]{trackchanges} 
\usepackage[tight,nice]{units}
\usepackage{url} 

\allowdisplaybreaks

\newcommand{\cites}[1]{\citeauthor{#1}'s~(\citeyear{#1})}

\journal{XXXXXXXXXXX}

\begin{document}
\begin{frontmatter}

\title{A generalized Townsend's theory for Paschen curves in planar, cylindrical, and spherical geometries}

\author[ERAU,FIT]{Jeremy A. Riousset}
\author[UO]{Joshua M\'endez Harper}
\author[UO]{Josef Dufek}
\author[ERAU]{Jacob A. Engle}
\author[ERAU]{Jared P. Nelson}
\author[FIT]{Annelisa B. Esparza}

\address[ERAU]{Department of Physical Sciences, Embry-Riddle Aeronautical University, Daytona Beach, FL, 32114, USA}
\address[FIT]{Aerospace, Physics \& Space Sciences Department, Florida Institute of Technology, Melbourne, FL, 32901, USA}
\address[UO]{Department of Earth Science, University of Oregon, Eugene OR, 97403, USA}

\begin{abstract}
	In this work, we focus on plasma discharges produced between two electrodes with a high potential difference, resulting in the ionization of the neutral particles supporting a current in the gaseous medium. At low currents and low temperatures, this process can create luminescent emissions: the so-called glow and corona discharges. The parallel plate geometry used in \cites{Townsend:1900} theory lets us develop a theoretical formalism, with explicit solutions for the critical voltage effectively reproducing experimental Paschen curves. However, most discharge processes occur in non-parallel plate geometries, such as discharges between grains or ice particles in multiphase flows. Here, we propose a generalization of the classic parallel plate configurations to concentric spherical and coaxial cylindrical geometries in Earth, Mars, Titan, and Venus atmospheres. In a spherical case, a small radius effectively represents a sharp tip rod, while larger, centimeter-scale radii represents rounded or blunted tips. Similarly, in a cylindrical case, a small radius would correspond to a thin wire. We solve continuity equations in the gap and estimate a critical radius and minimum breakdown voltage that allows ionization of neutral gas and formation of a glow discharge. We show that glow coron\ae~form more easily in Mars's low-pressure, $\rm CO_2$-rich atmosphere than in Earth's high-pressure atmosphere. Additionally, we present breakdown criteria for Titan and Venus. We further demonstrate that critical voltage minima occur at \unit[0.5]{cm$\cdot$Torr} for all three investigated geometries, suggesting easier initiation around millimeter-size particles in dust and water clouds and could be readily extended to examine other multiphase flows with inertial particles.
\end{abstract}

\end{frontmatter}

\section*{Plain Language Summary}
In this work, we focus on plasma discharges between two electrodes with a high voltage difference. The result is a conversion of the medium from a dielectric to a conductor. At low currents and low temperatures, this process can create luminescent emissions: the so-called glow and corona discharges. We extend the parallel plate geometry developed in \cites{Townsend:1900} classical theory to determine the critical discharge voltages of spheres and cylinders more likely to be encountered as particles in an atmosphere. Here, we propose a generalization of the classic parallel plate configurations to concentric spheres and coaxial cylinders in Earth, Mars, Venus, and Titan atmospheres. We computationally solve the continuity equations in the gap between objects and ultimately calculate critical electric fields for self-sustained discharges. We show that glow coron\ae~form more easily in Mars's low-pressure, $\rm CO_2$-rich atmosphere than in Earth's high-pressure atmosphere. Additionally, we present breakdown criteria for Titan and Venus. We further demonstrate that critical voltage minima occur near $\unit[0.5]{cm\cdot Torr}$ for all three investigated geometries, suggesting easier initiation around millimeter-size particles in dust and water clouds.

\section{Introduction}\label{sec:Introduction}

The recent and planned \textit{in-situ} exploration of planetary bodies in the solar system motivates a better understanding of electrostatic hazards under conditions relevant to each object. Specifically, the potential for discharge involves a complex interplay between atmospheric pressure variation, gas composition, realistic geometries of charged surfaces, and the presence of suspended solids in the atmosphere. The near-surface, diffuse conditions on present-day Mars may, in particular, present hazards associated with electrostatic discharges for both robotic endeavors and potential crewed missions \citep{Yair:2012}. Furthermore, the presence (or absence) of electrical discharges could have important implications for atmospheric chemistry and habitability \citep{Tennakone:2016, Hess:2021}. Any dielectric breakdown starts when the ambient electric field $E$ exceeds a threshold $E_{\rm th}$ \citep[e.g.,][p.~128]{Raizer:1991}, which depends on the nature of the discharge (e.g., leader, streamer, or glow) and its polarity \citep[see e.g.,][Figure~1 for discharge in air]{Pasko:2006}. 
Putative and confirmed extraterrestrial electrical discharges have been the topic of several studies \citep[see reviews by][and references therein]{Leblanc:2008,Riousset:2019}. While most investigations have focused on lightning as a ``transient, high-current electrical discharge whose path length is measured in kilometers'' \citep[p.~8~\&~Table~14.1]{Uman:2001}, a noteworthy few have also investigated Transient Luminous Events, TLEs \citep[e.g.,][]{Bering:2004,Dubrovin:2010,Yair:2012} and small-scale spark or glow discharges \citep[e.g.,][]{MendezHarper:2018, Mendez:2021a}. Elucidating discharge criteria on extraterrestrial environments is complicated by a profound dearth of \textit{in-situ} observational data. In the context of Mars, for example, the unfortunate fate of ExoMars' Schiaparelli module \citep{Deprez:2014} prevented the first direct measurements of the electric field at the surface of the planet. Insight into the atmospheric electrical environment on Venus and Titan, the two other rocky worlds in our solar systems with atmospheres thick enough to support gas breakdown, is also scant. Consequently, indirect measurements and analogies remain the only ways to gain insight into breakdown processes in planetary atmospheres. 

The diverse span of atmospheric conditions on worlds in our own solar system suggests that the criteria that lead to breakdown in extraterrestrial environments may be equally disparate. Although both Mars and Venus host $\rm~CO_2$-rich atmospheres, Venus maintains a near-surface atmospheric pressure $\sim10^4$ times higher than the Martian one \citep{Zasova:2007, Jakosky:2015, Jakosky:2015b, Sanchez:2017}. On Titan, the atmospheric surface pressure is only slightly higher than Earth's. However, Titan's atmosphere is 4 times denser than Earth's and significantly colder (\unit[90]{K} for Titan v. \unit[287]{K} for Earth \citep{Horst:2017}). Important chemical differences between worlds exist, too. Methane, for instance, is an important constituent of Titan's nitrogen-rich atmosphere. Oxygen, while abundant in Earth's atmosphere, exists in trace amounts or is absent in the atmospheres of the other three worlds. Likewise there is significant variability in the composition, abundance, and presence of particulates in these atmospheres (e.g. silicate dust, ice, hydrocarbons), and multiphase topologies may also be important for local discharge events. Using this diversity of atmospheric conditions (summarized in Table~\ref{tab:atm}), we revisit \cites{Townsend:1900} seminal model for self-sustained dielectrical breakdown between parallel electrodes. Townsend developed the theory supporting what is now known as \cites{Paschen:1889} law. Paschen's law states that the breakdown voltage between two electrodes is a function of the product of the pressure, $p$, and interelectrode distance, $d$. \cite{Townsend:1900} proved that this scaling law comes from the exponential increase of electron number density via avalanche multiplication and secondary ionization \citep[e.g.,][pp.~31--32]{Bazelyan:1998}. Interestingly, these early studies already involved experiments in air, carbon dioxide, and hydrogen. These gases contribute significantly to many planetary atmospheres in our solar system, demonstrating that discharge processes are highly dependent on gas composition. 

\begin{table}[!b]
\vspace*{-1em}
\centering
\begin{tabular}{cl@{}@{}cccc}
	\hline
	\hline
	& & Earth & Mars & Titan & Venus\\
	\hline
	\multirow{8}{*}{\rotatebox[origin=c]{90}{Molar fraction}}& $\rm Ar$ & $9.05\times10^{-3}$ & $1.60\times10^{-2}$ & $2.4\times10^{-2}$ & --\\
	& $\rm CH_4$ & -- & -- & $2.7\times10^{-2}$ & -- \\
	& $\rm CO$ & $1.84\times10^{-7}$ & -- & -- & -- \\
	& $\rm CO_2$ & $3.79\times10^{-3}$ & $95.7\times10^{-2}$ & -- & $96.2\times10^{-2}$\\
	& $\rm He$ & $5.04\times10^{-6}$ & -- & -- & --\\
	& $\rm N_2$ & $75.68\times10^{-2}$ & $2.7\times10^{-2}$ & $94.9\times10^{-2}$ & $3.5\times10^{-2}$\\
	& $\rm N_2O$ & $3.43\times10^{-7}$ & -- & -- & -- \\
	& $\rm O_2$ & $20.30\times10^{-2}$ & -- & -- & -- \\
	& $\rm O_3$ & $3.01\times10^{-8}$ & -- & -- & --\\
	\hline
	& $\unit[T]{(K)}$ & $273.04$ & $231.2$ & $93.9$ & $737$\\
	& $\unit[N]{(m^{-3})}$ & $2.688\times10^{25}$ & $1.889\times10^{23}$ & $1.150\times10^{26}$ & $9.131\times10^{26}$\\
	\hline
	\multirow{4}{*}{\rotatebox[origin=c]{90}{Coeff.}} & $\unit[A]{(10^{-20}m^2)}$ & $1.04$ & $2.11$ & $2.14$ & $1.42$\\
	& $\unit[B]{(Td^{-1})}$ & $596.8$ & $594.3$ & $602.5$ & $723.4$ \\
	& $\unit[C]{(10^{24}/(Vms))}$ & $3.35$ & $12.32$ & $12.38$ & $3.75$\\
	& $\unit[D]{}$ & $-0.23$ & $-0.46$ & $-0.46$ & $-0.23$ \\
	\hline
	\hline
\end{tabular}
\caption{Input parameters for BOLSIG runs. Atmospheric parameters are from NASA's Global Reference Atmospheric Models (GRAMs; EarthGRAM by \cite{Leslie:2008},  MarsGRAM by \cite{Justh:2010}, TitanGRAM by \cite{Justh:2020}, and  VenusGRAM by \cite{Justh:2021}) taken at the surface $z$=\unit[0]{km} on January \nth{1}, 2000, 1200 UT, at $0^\circ$ latitude and $0^\circ$ longitude. These are the same surface conditions as in \citep{Riousset:2019}. The coefficients $A$, $B$, $C$, and $D$ define $\tilde{a}/N$ and $\tilde{\mu}\times N$ in \eqref{eq:Fits_N}.}
\label{tab:atm}
\end{table}

The elegance of Townsend's theory rests in its simplicity and the sole requirement of an exponential approximation for the effective ionization coefficient $\alpha$. We revisit Townsend's theory from first principles in Section~\ref{sec:Model}. Townsend's theory, however, assumes that the discharge occurs between two infinite parallel plates (i.e., a 1-D Cartesian geometry). To approach this configuration, experimental setups have adopted large flat electrodes with large radii $R$, and small gap size, $d$, so that $R\gg d$ (\citealp[e.g.,][p.~53]{Raizer:1991}; \citealp{Lowke:2003, Stumbo:2013}). While such configurations are suitable for laboratory experiments, they may not be representative of real discharge processes that invariably deal with complex geometries. In fact, natural electrical discharge events are almost always associated with multiphase flows. For instance, discharges on Mars may occur between small sand grains. Similarly, arcing could occur between two voltage-carrying conductors under appropriate pressure-distance products. Thus, in the remainder of Section ~\ref{sec:Model}, we demonstrate that an extension to cylindrical and spherical geometries is possible for Townsend's theory provided one approximates the mobility $\mu$. We further develop a generalized Townsend's criterion for the ignition of self-sustained gas discharges for coaxial cylinders and concentric spheres. We show that the numerical solutions of these equations yield the critical potential $V_{\rm cr}$ and corresponding electric field $E_{\rm cr}$ and satisfy the same similarity laws as first introduced by \cite{Paschen:1889}. Sections~\ref{sec:Results} and~\ref{sec:Discussion} will respectively discuss the results and implications of the new formalism, while section~\ref{sec:Conclusions} will summarize the principal contributions of this paper.

\section{Model Formulation}\label{sec:Model}
\begin{figure}[t]
\centering
\includegraphics[width=\textwidth]{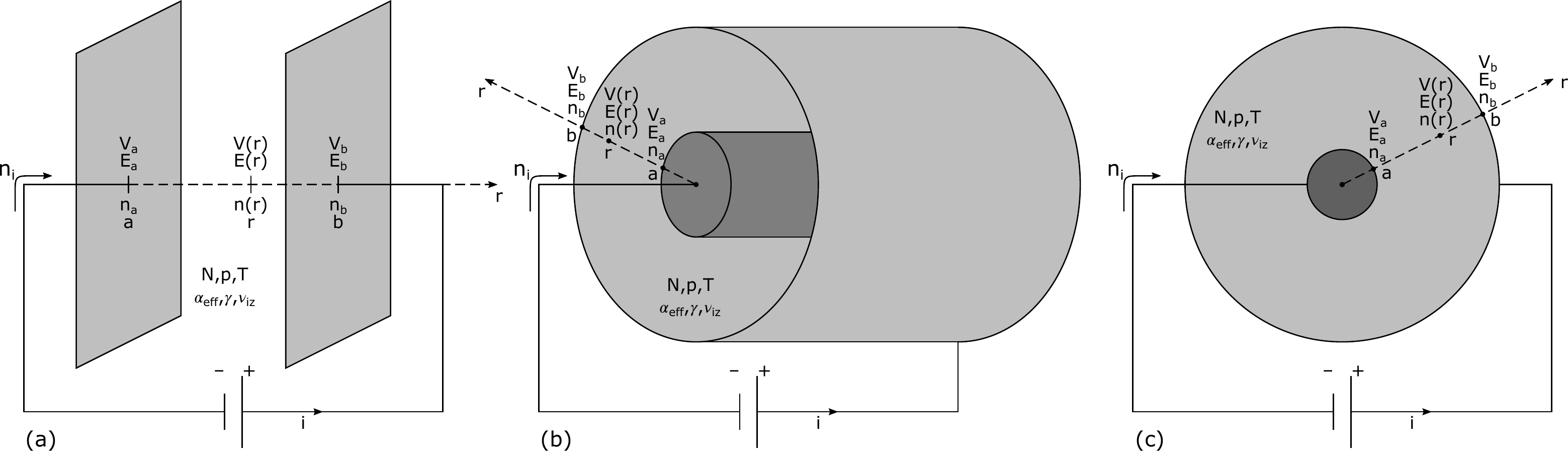}
\caption{Townsend's discharge in one-dimensional geometries: (a) Parallel plates (Cartesian); (b) Coaxial cylindrical electrodes; (c) Concentric spherical electrodes. The gas in between the electrodes has the number density $N~\unit{(m^{-3})}$ at the temperature $T~\unit{(K)}$ under the pressure $p~\unit{(Pa)}$. The avalanche is characterized by Townsend's effective ionization coefficient $\alpha~\unit{(m^{-1})}$, the secondary ionization coefficient $\gamma$, and effective ionization frequency $\nu_{\rm i}~\unit{(s^{-1})}$. The quantities $n_i$, $n_a$, $n(r)$, and $n_b$ correspond to the electron density in $\unit{m^{-3}}$ carried by the electronic current $i$, emitted from the cathode at $a$, measured at $r$, and received at the anode at $b$, respectively ($a\leq r\leq b$). The corresponding electric potential and field are denoted $V~\unit{(V)}$ and $E~\unit{(V/m)}$.}
\label{fig:Geometries}
\end{figure}

This section describes the model used to develop a criterion for the initiation of self-sustained glow discharge between two one-dimensional electrodes located at $r$=$a$ and $b$, where $r$ is a coordinate along the direction normal to the surface of the electrode (Figure~\ref{fig:Geometries}). 

In the absence of free electric charges, Gauss's law for electric field $\va*E$ reduces to $\divergence{\va*{E}}=0$. It further simplifies into:
\begin{linenomath*}
\begin{align}
\frac{1}{r^\delta}\dv{r^\delta E(r)}{r} = 0, \label{eq:Gauss1D}
\end{align}
\end{linenomath*}
where $\delta=0$, $1$, and $2$ for the Cartesian, cylindrical, spherical 1-D geometries displayed in Figures~\ref{fig:Geometries}a, \ref{fig:Geometries}b, and \ref{fig:Geometries}c, respectively. If {\em space charges do not contribute significantly to the total electric field between the electrodes}, then:
\begin{linenomath*}
\begin{align}
E(r)=E_a\left(\dfrac{a}{r}\right)^\delta,\label{eq:E}
\end{align}
\end{linenomath*}
with $E_a=E(a)$ and $a\leq r\leq b$.

\begin{figure}[!b]
\centering
\includegraphics[width=30pc]{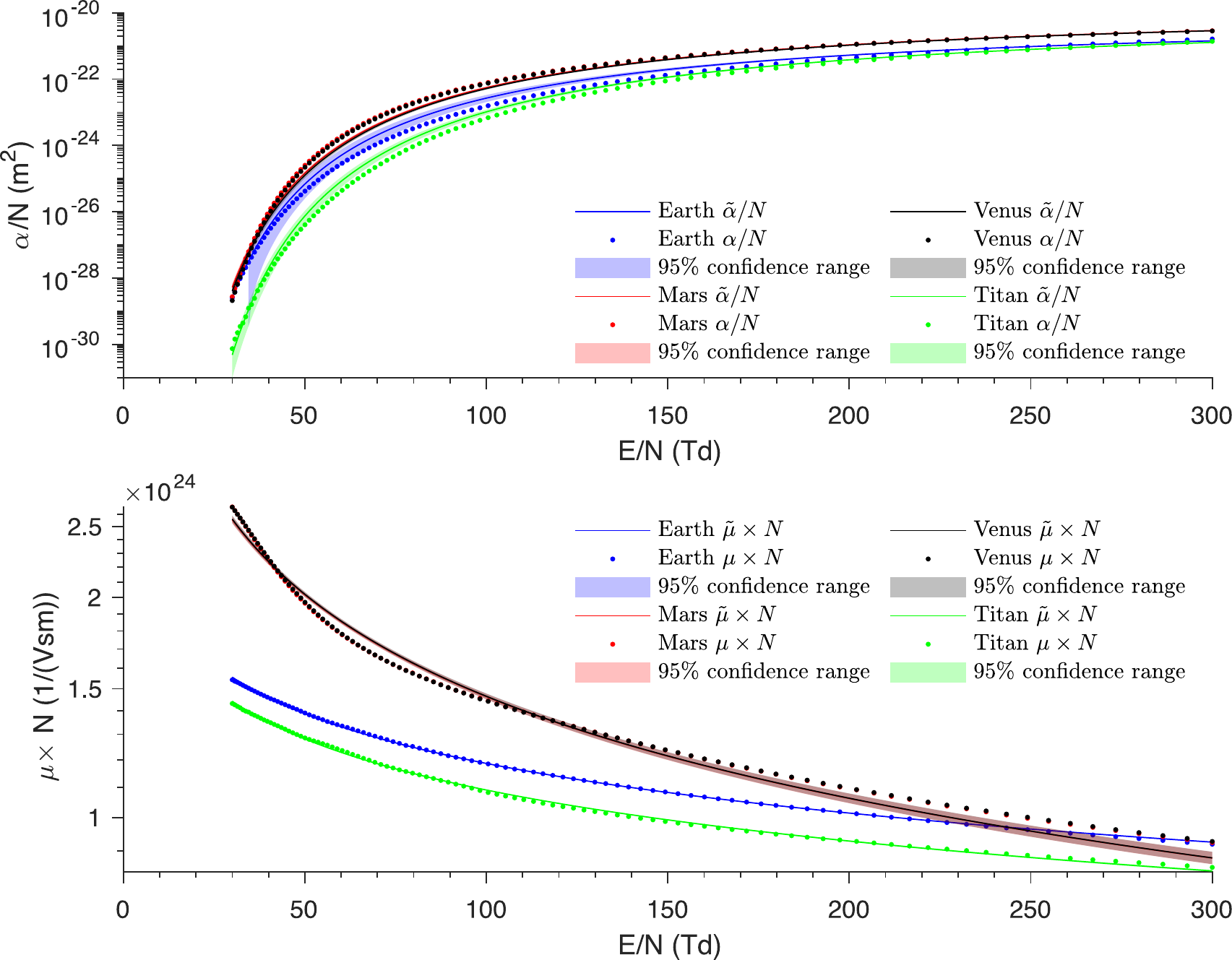}
\caption{Scaling laws for (a) the reduced effective Townsend's ionization coefficient $\alpha/N$ and (b) reduced mobility $\mu\times N$ plotted against the reduced electric field $E/N$. Blue, red, gray, and green colors correspond to Earth-, Mars-, Venus-, and Titan-like atmospheres, respectively (see Table~\ref{tab:atm}).}
\label{fig:Scaling}
\end{figure}

The ignition of an electron avalanche between two electrodes depends on the effective ionization frequency $\nu_{\rm i}$ and the poorly understood secondary ionization coefficient $\gamma$ \citep[p.~74]{Raizer:1991}. Townsend's effective ionization coefficient $\alpha$ provides a convenient description of the primary ionization per unit length:
\begin{linenomath*}
\begin{align}
\alpha=\dfrac{\nu_{\rm i}}{\norm{\va*u}}=\dfrac{\nu_{\rm i}}{u}
\label{eq:aa_u}
\end{align}
\end{linenomath*} 
where the drift velocity $\va*u$ depends on the mobility $\mu$ as follows \citep[e.g.,][p.~66]{Chen:1984}:
\begin{linenomath*}
\begin{align}
\va*u=\mu(E)\va*{E}.
\label{eq:u_E}
\end{align}
\end{linenomath*} 
Thus, $\alpha$ depends on $E$ as follows:
\begin{linenomath*}
\begin{align}
\alpha(E)=\dfrac{\nu_{\rm i}(E)}{\mu(E)E}
\label{eq:aa}
\end{align}
\end{linenomath*} 
Townsend's theory provides an analytical solution to Paschen curves if $\alpha$ approximately fits an exponential function:
\begin{linenomath*}
\begin{align}
\tilde{\alpha} = Ap\exp(-Bp/E),
\label{eq:aa_p}
\end{align}
\end{linenomath*} 
where $p$ is the neutral gas pressure \citep[e.g.,][pp.~149]{Raizer:1991}. Experimental studies typically adopt pressure-based scaling with $p$ in \unit{Torr} and $\alpha$ in $\unit{cm^{-1}}$ \citep[e.g.,][pp.~133]{Raizer:1991} giving $\alpha/p$ in $\unit{1/(cm\cdot Torr)}$. On the other hand, theoretical investigations usually prefer density-based scaling with $N$, the neutral gas number density in $\unit{m^{-3}}$ and $\alpha$ in $\unit{m^{-1}}$, returning $\alpha/N$ in $\unit{m^2}$ \citep[e.g.,][p.~545]{Hagelaar:2015,Lieberman:2005}. Both formulations are equivalent, provided that the system remains approximately at the temperature $T$ and that the gas obeys the ideal gas law, namely $p=Nk_{\rm B}T$, where $k_{\rm B}$ is the Boltzmann constant. Consequently, we can write:
\begin{linenomath*}
\begin{subequations}\label{eq:Fits_N}
	\begin{align}
	\dfrac{\tilde{\alpha}}{N} &= A\exp(-\dfrac{B}{\nicefrac{E}{N}})\label{eq:aa_N}
	\end{align}
	\begin{align}
	\tilde{\mu}\times N &= C\left(\dfrac{E}{N}\right)^D
	\label{eq:muxN}
	\end{align}
\end{subequations}
\end{linenomath*}
where $A$, $B$, $C$, and $D$ are the coefficients from a fit to the reduced Townsend's effective ionization $\alpha/N$ and mobility $\mu\cp N$ (Figure~\ref{fig:Scaling}) for the atmospheres considered here (Table~\ref{tab:atm}).

The condition for self-sustainability of Townsend's discharges in any of the geometries shown in Figure~\ref{fig:Geometries} starts with the continuity equation:
\begin{linenomath*}
\begin{align}
\pdv{n}{t} + \divergence{n\va*u} &= n\nu_{\rm i} \label{eq:Continuity}
\end{align}
\end{linenomath*}
where $n$ is the plasma density. 

Combining Equations~\eqref{eq:Gauss1D}, \eqref{eq:aa_u}, and \eqref{eq:Continuity} in a steady state ($\pdv*{}{t}=0$) yields $\frac{1}{r^\delta}\dv{}{r}\left(r^\delta nu\right) = n\alpha u$. Using Equation~\eqref{eq:u_E}, this simplifies further into:
\begin{linenomath*}
\begin{align}
\dv{\ln(r^\delta n\mu(E)E)}{r}&=\alpha(E).\label{eq:ContODE}
\end{align}
\end{linenomath*}
From Equation~\eqref{eq:E}, we have $E_aa^\delta=E_bb^\delta$. If $A_{\rm av}=n_b/n_a$ is the avalanche coefficient defined as the ratio of the number densities $n_a=n(a)$ and $n_b=n(b)$, then Equation~\eqref{eq:ContODE} yields:
\begin{linenomath*}
\begin{align}
A_{\rm av}=\dfrac{ n_b}{n_a}&=\dfrac{ \mu(E_a)}{\mu(E_b)}\exp(\displaystyle\int_a^b\alpha(E)\dd{r})\label{eq:Cont_int}
\end{align}
\end{linenomath*}

\subsection{Sustainability}
Call $n_i$ the number density of charges from the electronic current $i$ and $n_\gamma$ the one from secondary avalanches between the electrodes, then the conservation of charge produces the system below:
\begin{linenomath*}
\begin{align}
\begin{cases}
n_a &= n_i + n_\gamma \\
n_\gamma &= \gamma(n_b-n_a) \\
n_b & = A_{\rm av}n_a \end{cases}.
\end{align}
\end{linenomath*}
An electron avalanche occurs when the ratio $n_b/n_i$ diverges \citep[e.g.,][Sec.~2.5]{Naidu:2013}. Then, the condition for initiating a self-sustained discharge follows from Equation~\eqref{eq:Continuity}: 
\begin{linenomath*}
\begin{align}
\dfrac{n_b}{n_i}&=\dfrac{\dfrac{A_{\rm av}}{\gamma}}{1+\dfrac{1}{\gamma}-A_{\rm av}}\rightarrow\infty,
\end{align}
\end{linenomath*}
which is satisfied when:
\begin{linenomath*}
\begin{align}
A_{\rm av}&=1+\dfrac{1}{\gamma}.\label{eq:SS}
\end{align}
\end{linenomath*}
Using Equation~\eqref{eq:SS} to substitute $A_{\rm av}$ in Equation~\eqref{eq:Cont_int} yields after simplifications:\footnote{Note that if $E_a=E_b$ (e.g., in a parallel plate geometry), one straightforwardly retrieves the classic formula \citep[e.g.,][p.~177]{Raizer:1991}.}
\begin{linenomath*}
\begin{align}   \Aboxed{\int_a^b\alpha(E)\dd{r}+\ln\left(\dfrac{\mu(E_a)}{\mu(E_b)}\right)&=\ln\left(1+\dfrac{1}{\gamma}\right)}. \label{eq:SS2}
\end{align}
\end{linenomath*}

In all three 1-D cases, Equations \eqref{eq:E}, \eqref{eq:aa_N}, and \eqref{eq:muxN} let us approximate $\alpha/N$ and $\mu\times N$ as a function of $a$, $r$, and $E_a$. Thus, the condition of self-sustainability Equation~\eqref{eq:SS2} in the absence of space charges and displacement field becomes:
\begin{linenomath*}
\begin{align}
\int_a^bAN\exp(-\dfrac{B}{\nicefrac{E_a}{N}}\left(\dfrac{r}{a}\right)^\delta)\dd{r}+D\ln\left(\left(\dfrac{b}{a}\right)^\delta\right)&=\ln\left(1+\dfrac{1}{\gamma}\right) \label{eq:SS3}
\end{align}
\end{linenomath*}
If $A$ and $B$ are converted to $\unit{1/(cm\cdot Torr)}$ and $\unit{V/(cm\cdot Torr)}$ and $d$ is the distance between the electrodes ($b=a+d$), then we can show that the scalability of $E_a/p$ naturally derives from Equation~\eqref{eq:SS3} as follows:
\begin{linenomath*}
\begin{align}
\int_0^dAp\exp(-\dfrac{B}{\nicefrac{E_a}{p}}\left(1+\dfrac{pr}{pa}\right)^\delta)\dd{r}+D\ln\left(\left(1+\dfrac{pd}{pa}\right)^\delta\right)&=\ln\left(1+\dfrac{1}{\gamma}\right) \label{eq:SS4}
\end{align}
\end{linenomath*}
The critical electric field $E_{\rm cr}$ is measured at $r=a$, therefore $E_{\rm cr}=\abs{E_a}$. Consequently, Equation~\eqref{eq:SS4} yields the following results for the specific geometries described in Figure~\ref{fig:Geometries}:
\begin{linenomath*}
\begin{subequations}\label{eq:Ecr}
	\footnotesize
	\begin{align}
	\exp\left(-\dfrac{Bp}{E_{\rm cr}}\right) &=\dfrac{1}{Apd}\ln\left(1+\dfrac{1}{\gamma}\right) & \delta&=0\label{eq:Ecr_Ca}\\
	-\dfrac{E_{\rm cr}}{Bp}\left[\exp\left(-\dfrac{Bp}{E_{\rm cr}}\left(1+\dfrac{pd}{pa}\right)\right)-\exp\left(-\dfrac{Bp}{E_{\rm cr}}\right)\right]&=\dfrac{1}{A p a}\ln\left(\dfrac{\left(1+\dfrac{1}{\gamma}\right)}{\left(1+\dfrac{pd}{pa}\right)^D}\right) & \delta&=1\label{eq:Ecr_Cy}\\
	\sqrt{\dfrac{E_{\rm cr}}{Bp}}\left[\erf\left(\sqrt{\dfrac{E_{\rm cr}}{Bp}}\left(1+\dfrac{pd}{pa}\right)\right) -\erf\left(\sqrt{\dfrac{E_{\rm cr}}{Bp}}\right)\right]&=\dfrac{2}{\sqrt{\pi}}\dfrac{1}{A p a}\ln\left(\dfrac{\left(1+\dfrac{1}{\gamma}\right)}{\left(1+\dfrac{pd}{pa}\right)^{2D}}\right) &\delta&=2\label{eq:Ecr_Sp}
	\end{align}
\end{subequations}
\end{linenomath*}
where $\erf(x)=\displaystyle\dfrac{2}{\sqrt{\pi}}\int_0^xe^{-t^2}\dd{t}$ is the Gauss error function \citep[e.g.,][p.~203]{Lipschutz:2018}.

\subsection{Critical voltage}
Paschen curves are plots of the product pressure times density $pd$ versus critical electric potential $V_{\rm cr}$. This potential measured between the electrodes at $a$ and $b$ ($V_{\rm cr}=V_b-V_a$) corresponds to the voltage necessary to initiate a self-sustained discharge and obeys the classic definition $V(r)=-\int_a^b\va*{E}\cdot\dd{\va*{r}}$ \citep[e.g.,][p.~62]{Zangwill:2019}. Equation~\eqref{eq:E} then yields $V_{\rm cr}$ for the three cases of Figure~\ref{fig:Geometries}:
\begin{linenomath*}
\begin{subequations}\label{eq:Vcr}
	\begin{align}
	V_{\rm cr}&=E_{\rm cr}d & \delta&=0\label{eq:Vcr_Ca}\\
	V_{\rm cr}&=E_{\rm cr}a\cdot\ln\left(1+\dfrac{d}{a}\right) & \delta&=1\label{eq:Vcr_Cy}\\
	V_{\rm cr}&=E_{\rm cr}d\cdot\left(1+\dfrac{d}{a}\right)^{-1} & \delta&=2\label{eq:Vcr_Sp}
	\end{align}
\end{subequations}
\end{linenomath*}

For $\delta=0$ (case of parallel plates), Equations~\eqref{eq:Ecr_Ca} and \eqref{eq:Vcr_Ca} simplify into the well-established formula \citep[e.g.,][p.~133]{Raizer:1991}:
\begin{linenomath*}
\begin{align}
V_{\rm cr}&=\dfrac{Bpd}{\ln\left(\dfrac{Apd}{\ln\left(1+\dfrac{1}{\gamma}\right)}\right)}
\end{align}
\end{linenomath*}
If Equations \eqref{eq:Ecr_Cy} and \eqref{eq:Ecr_Sp} had known analytical solutions, one could straightforwardly obtain solutions in the cylindrical and spherical geometries ($\delta$=1 and 2, respectively) from Equations \eqref{eq:Vcr_Cy} and \eqref{eq:Vcr_Sp}. In the absence of closed-form solutions, we use MATLAB \texttt{fzero} root-finding algorithm to numerically solve Equation \eqref{eq:Ecr} for $E_{\rm cr}$ given specific values of $pa$ and $pd$. This function combines the bisection, secant, and inverse quadratic interpolation methods and is based on the works by \cite{Brent:1973} and \cite{Forsythe:1976}. We use the values of $E_{\rm cr}$ to deduce the critical voltage $V_{\rm cr}$ from Equation \eqref{eq:Vcr} given $pa$, $pd$, and $\delta$. 

In the next section, we present the results from our calculation as surface plots for all three geometries of Figure~\ref{fig:Geometries} for the environments described in Table~\ref{tab:atm}. We also compare the results to experimental data from the peer-reviewed literature. 

\section{Results} \label{sec:Results}
\begin{sidewaysfigure}
\centering
\includegraphics[width=\textwidth]{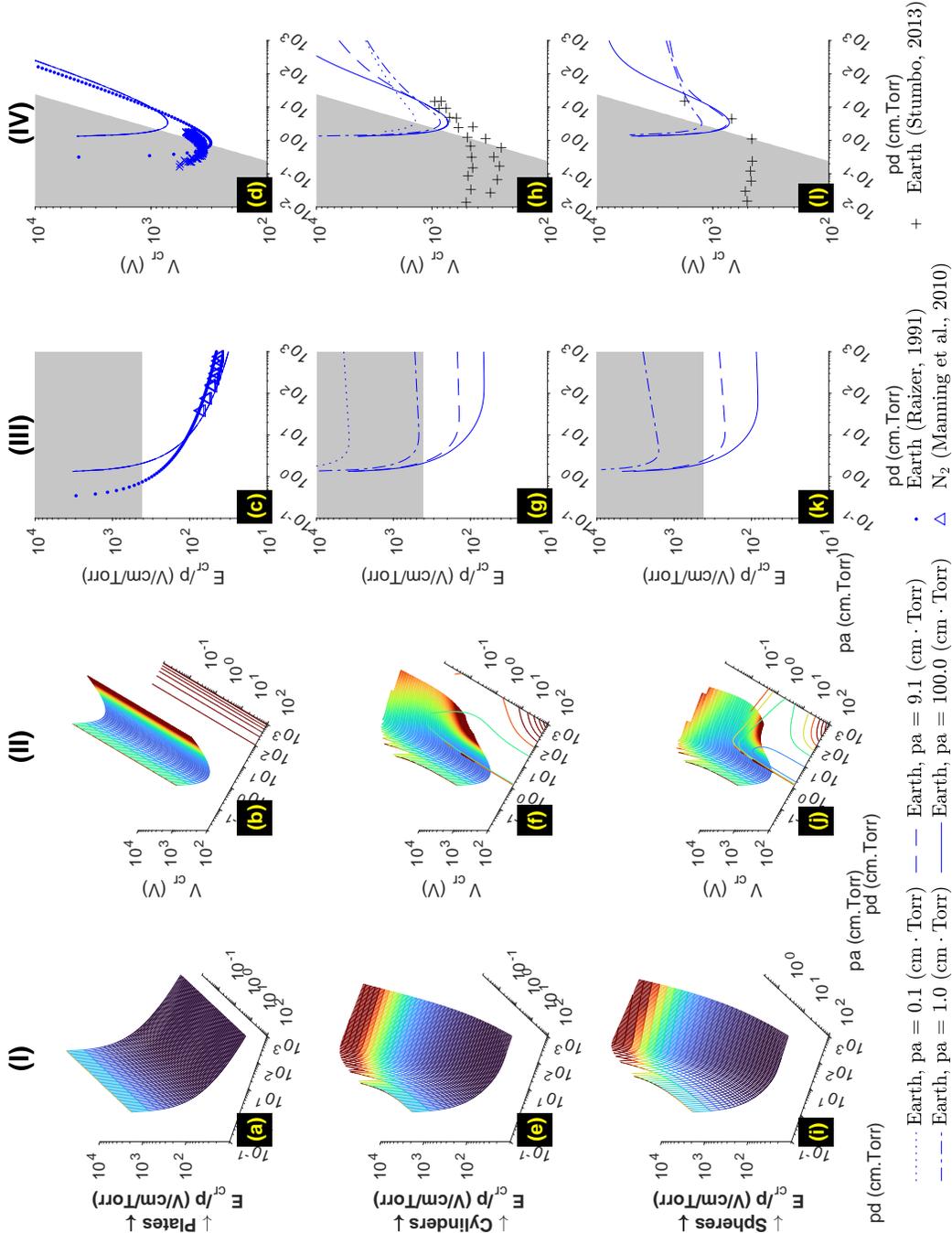}
\caption{Breakdown criteria in the Earth-like atmosphere described in Table~\ref{tab:atm}. The reduced critical electric field $E_{\rm cr}/p$ is displayed as a function of the reduced size of electrode $pa$ and distance $pd$ between electrodes $a$ and $b$ in column (I). Column (III) shows $E_{\rm cr}/p$ vs. $pd$ for selected $pa$-values for comparison with experimental data. Columns (II) and (IV) are the same as columns (I) and (III) but for the critical voltage $V_{\rm cr}$. The first, second, and third rows displays the results for electrodes with the following geometries: parallel plates (a-d), coaxial cylinders (e-h), and concentric spheres (i-l). The shaded areas correspond to domains where $E\geq10E_{\rm k}$.}
\label{fig:Earth}
\end{sidewaysfigure}

\begin{sidewaysfigure}
\centering
\includegraphics[width=\textwidth]{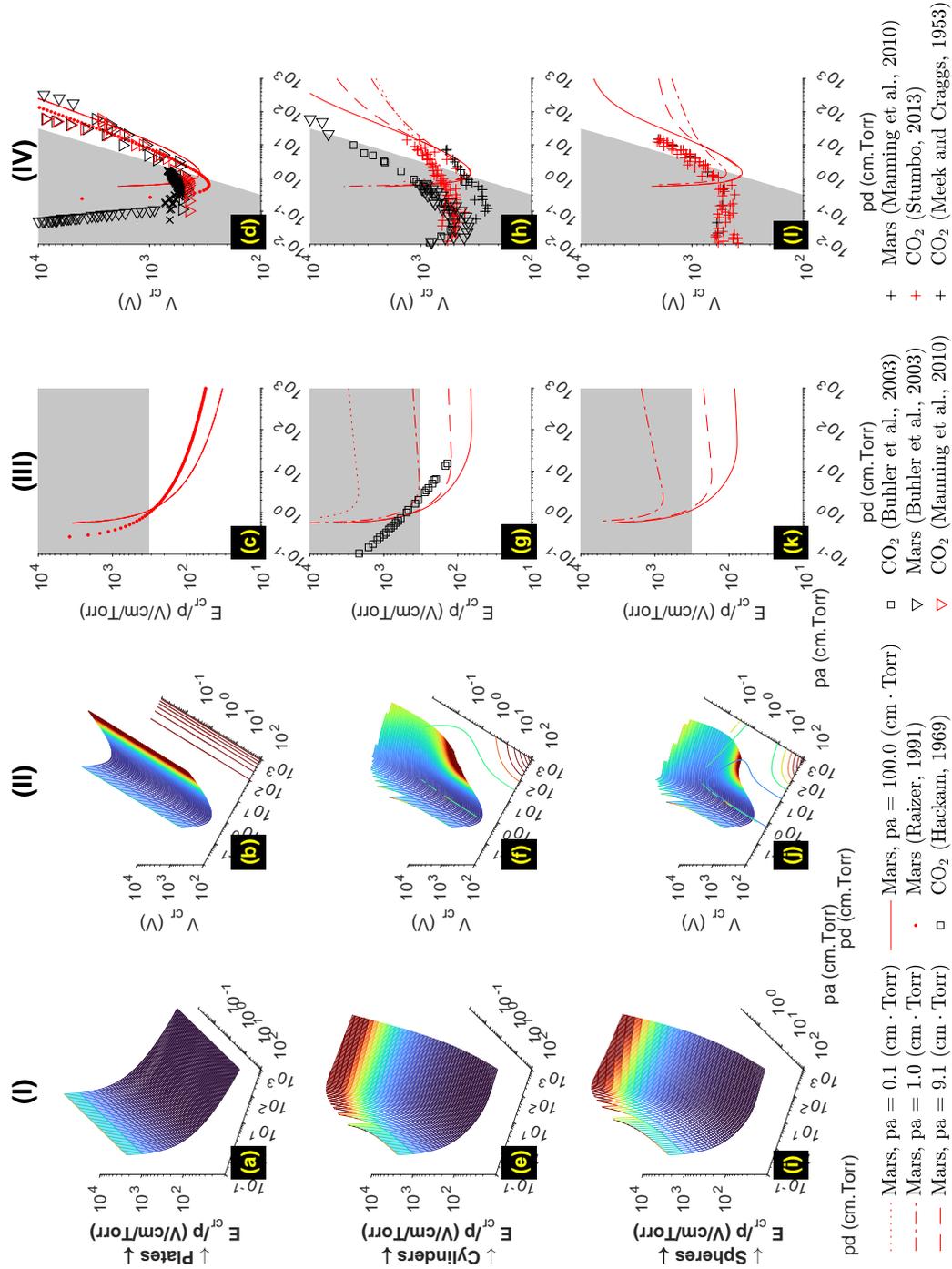}
\caption{Same as Figure~\ref{fig:Earth} for the Mars-like environment described in Table~\ref{tab:atm}.}
\label{fig:Mars}
\end{sidewaysfigure}

\begin{sidewaysfigure}
\centering
\includegraphics[width=\textwidth]{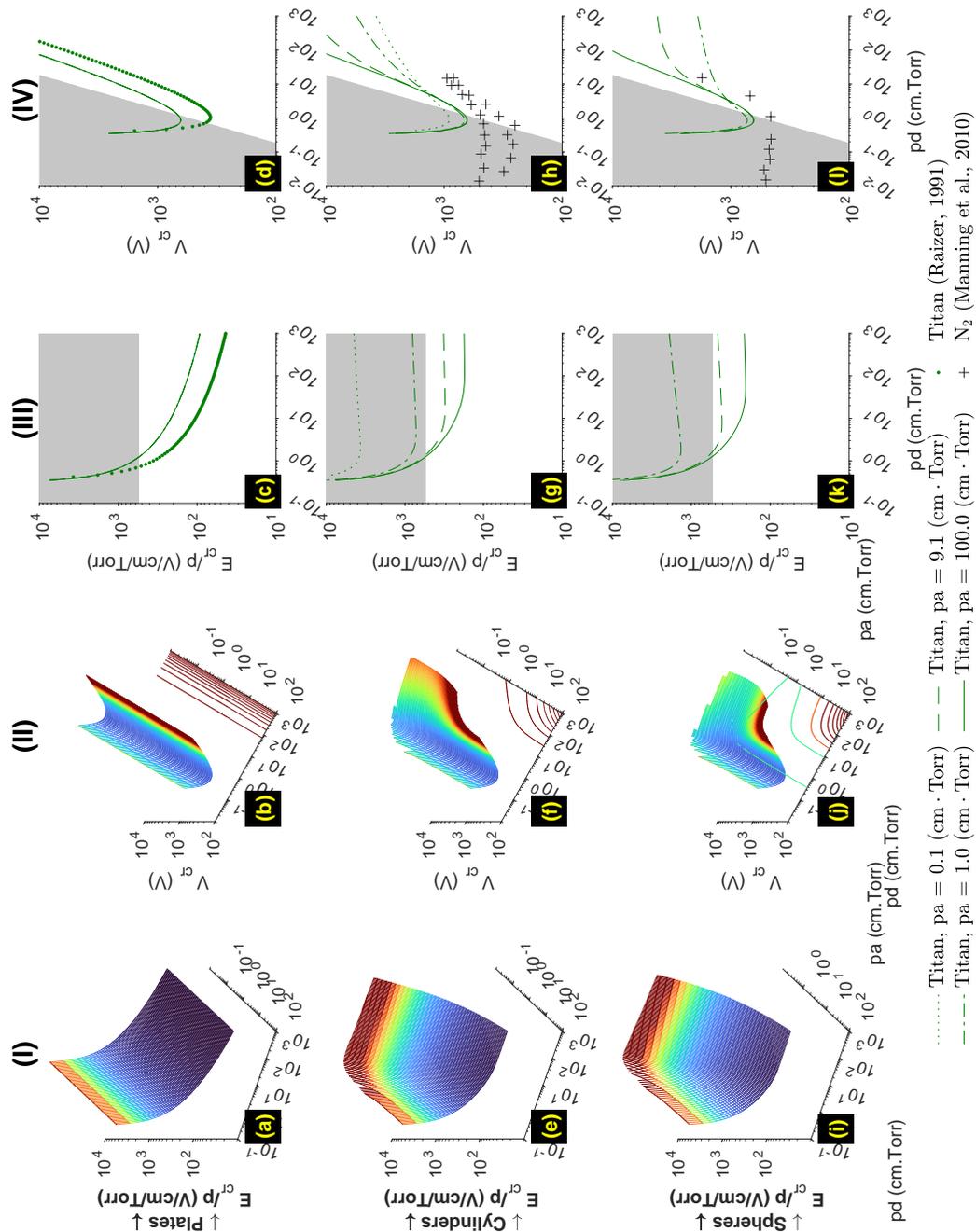}
\caption{Same as Figure~\ref{fig:Earth} for the Titan-like environment described in Table~\ref{tab:atm}.}
\label{fig:Titan}
\end{sidewaysfigure}

\begin{sidewaysfigure}
\centering
\includegraphics[width=\textwidth]{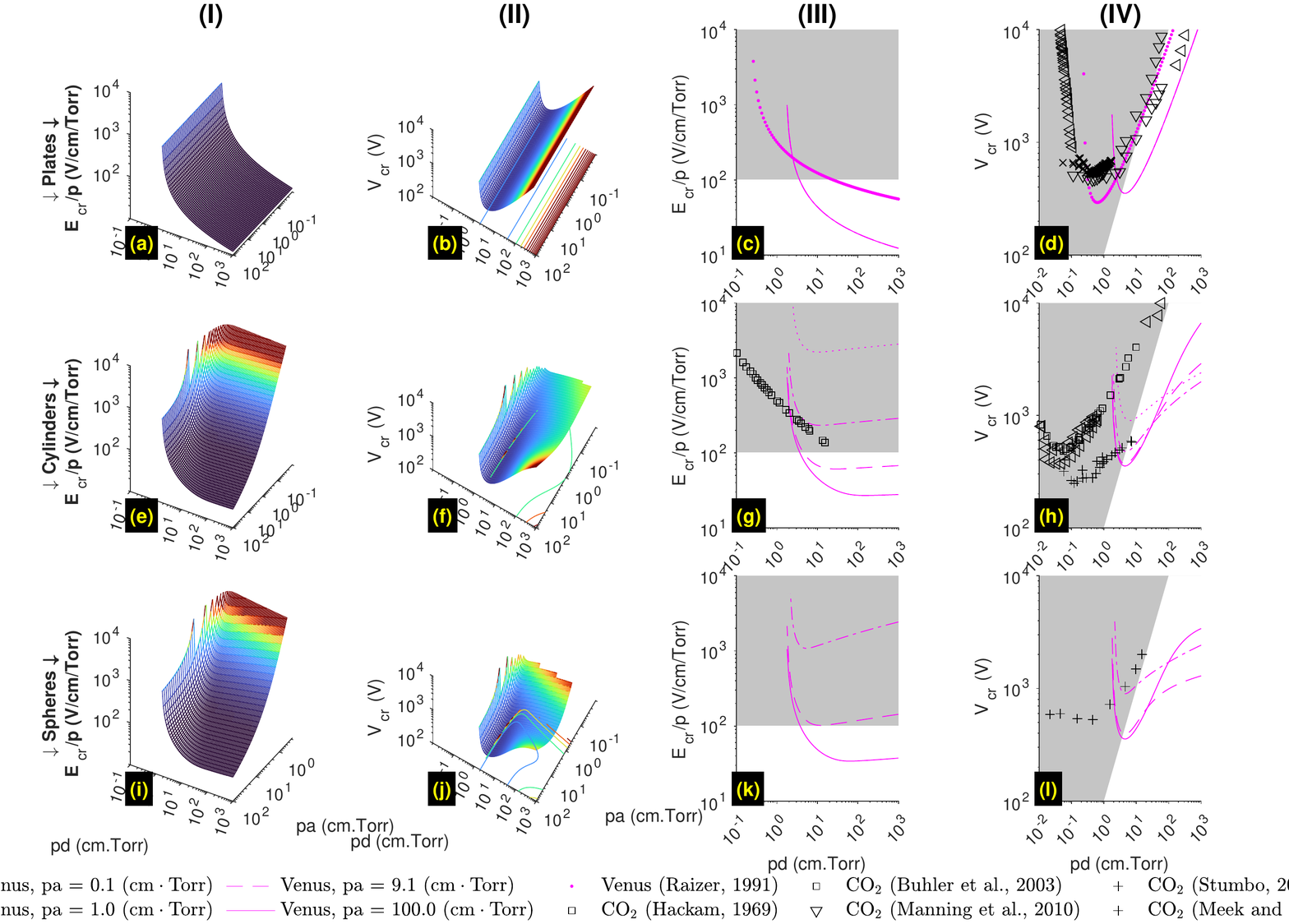}
\caption{Same as Figure~\ref{fig:Earth} for the Venus-like environment described in Table~\ref{tab:atm}.}
\label{fig:Venus}
\end{sidewaysfigure}

The near-surface atmospheric breakdown criteria for Earth, Mars, Titan, and Venus are summarized in Figures~\ref{fig:Earth} through \ref{fig:Venus}. In each figure, columns (I) and (II) respectively display the critical electric field $E_{\rm cr}$ and potential $V_{\rm cr}$ for the various geometries explored here as functions $pd$ and $pa$. The results are displayed for values of $pa$ from $10^{-1}$ to $\unit[10^{+3}]{cm\cdot Torr}$ and $pd$ from $10^{-1}$ to $\unit[10^{+3}]{cm\cdot Torr}$. The use of pressure-scaled dimensions eases the comparison with experimental data in columns (III) and (IV). Therefore, Figures~\ref{fig:Earth} to \ref{fig:Venus} use pressure-scaled values ($E/p$, $pa$, $pd$) rather than number-density scaled parameters (e.g., $E/N$, $\alpha/N$, $\mu\times N$ in Figure~\ref{fig:Scaling}). The conversion is possible using the neutral temperatures given in Table~\ref{tab:atm} and the ideal gas law discussed in Section~\ref{sec:Model} (see \ref{app:Conversions} for details).

In Figures~\ref{fig:Earth} to \ref{fig:Venus}, the first, second, and third rows display the results for parallel plates, coaxial cylinders, and concentric spherical electrodes, respectively. Specifically, panels (a) and (b) show the calculated values of $E_{\rm cr}$ and $V_{\rm cr}$ for the parallel plate geometry and confirm that the critical electric field and potential do not vary as a function of $pa$. Therefore, the surface plots effectively collapse into the well-known curves of Townsend's theory. For coaxial cylinders, panels (e) and (f), and concentric spheres, panels (i) and (j), Figures~\ref{fig:Earth} to \ref{fig:Venus} exhibit a similar dependence with $pd$, related to the separation between the electrodes, but also introduce a new dependence on $pa$, emphasizing the role of the size of the system for the initiation of self-sustained glow discharge. Conventional Paschen curves have a well-defined minimum, Stoletov's point, with a potential $V_{\rm min}=\dfrac{eB}{A}\ln(1+\dfrac{1}{\gamma})$ at $pd_{\min}=\dfrac{e}{A}\ln(1+\dfrac{1}{\gamma})$ \citep[e.g.,][p.~134]{Raizer:1991}. However, this minimum point becomes a minimum curve in cylindrical and spherical geometries (panels (f) and (j) in Figures~\ref{fig:Earth} to~\ref{fig:Venus}). As expected, the minimum of the surface plot for the parallel-plate case is independent of the value of $pa$ and obeys Stoletov's equations for $pd_{\min}$ and $V_{\min}$.

We compare our numerical results with other numerical calculations and experimental data in columns (III) and (IV) of Figures~\ref{fig:Earth} to \ref{fig:Venus}. There, we plot selected cross-sections from the surface plots in columns (I) and (II). Columns (III) and (IV) display $E_{\rm cr}/p$ and $V_{\rm cr}$, respectively, as a function of the parameter $pd$ for fixed values of $pa$: \unit[0.1, 1, $\sim$10, and 100]{cm$\cdot$Torr}. Experimental measurements in air are  rendered in blue markers: $\bullet$ and $\bigtriangleup$ for measurements from \citep{Raizer:1991} and $\times$ for \cites{Stumbo:2013} data. Red markers display data for Mars: $\bigtriangledown$ for \cites{Raizer:1991} and $+$ for \cites{Manning:2010} works. We further show data for pure $\rm CO_2$ with black markers where $\bullet$, $\square$, $\bigtriangledown$, $+$, $\times$, and $\vartriangleleft$ show the results from \citep{Raizer:1991}, \citep{Hackam:1969a}, \citep{Buhler:2003}, \citep{Manning:2010}, \citep{Stumbo:2013}, and \citep{Meek:1953}, respectively. For comparisons with Earth and Titan scenarios, we included experimental results  in $\rm N_2$ from \citep{Manning:2010}. 

As expected, self-sustained discharges between parallel plates do not depend on the parameter $pa$ (see panels (a) and (b) across Figures~\ref{fig:Earth} through ~\ref{fig:Venus}). Interestingly, Stoletov's points visible in column (IV) occur at similar values of $pd$ and $V_{\min{}}$ ($\approx \unit[0.1-1]{cm\cdot Torr}$ and $\approx\unit[2]{V/cm /Torr}$, respectively) for the various gas mixtures and geometries. Columns (III) and (IV) also show that the theoretical values from Section~\ref{sec:Model} underestimate experimental values of $E_{\rm cr}$ and $V_{\rm cr}$ for all geometries and values of $pa$. For a given system, if one splits the Paschen curve around the Stoletov's point, one can define a high electric field regime for $pd\ll pd_{\min{}}\approx\unit[0.1-1]{cm\cdot Torr}$ forming the left-hand branch of the curves and a high-pressure regime on the right-hand branch of curves for $pd\gtrsim pd_{\min{}}$. 


The right branches of the curves show promising agreement (within an order of magnitude) between theoretical curves and measurements provided that: (1) one accounts for the uncertainty in deriving the coefficients $A$, $B$, and $D$; and (2) one carefully considers the values of $pa$ best representing the geometries of concentric or coaxial electrodes. For these reasons, we shall note that the theoretical slopes closely follow those formed by the experimental measurements from various authors (see legends of Figures~\ref{fig:Earth} through \ref{fig:Venus} for details). The curves using $A$ and $B$ from \citep[p.~56 and `$\bullet$' markers in columns (III) and (IV)]{Raizer:1991} show the influence of these coefficients on the location of Stoletov's points in theoretical plots. While some authors have derived these values directly from the Paschen curves, we derived $A$, $B$, and $D$ from solutions to the Boltzmann equation (see Figure~\ref{fig:Scaling} and Section~\ref{sec:Model}) to maintain consistency of methodology across coefficients and gases. Another explanation of the aforementioned difference stems from the differences between pure gases and complex atmospheres (e.g., pure $\rm CO_2$ vs. Mars atmospheres, pure $\rm N_2$ vs air). Even in cases when the atmospheric composition is almost pristine (e.g., Mars's atmosphere is $\gtrsim95\%$ $\rm CO_2$), the presence of minor components can dramatically alter the condition for discharge initiation as evidenced by \cite{Riousset:2019} in the case of conventional breakdown $E_{\rm k}$.



\section{Discussion} \label{sec:Discussion}
The results presented in Section~\ref{sec:Results} differ from previous attempts at generalizing Townsend's theory of Paschen curves mainly in their full treatment of electron mobility in the continuity equation. Neglecting the volume increase along the avalanche path, \citet[p.~177]{Raizer:1991} or \citet[p.~100]{Meek:1953} straightforwardly rewrote the condition for initiation of self-sustained discharges Equation~\eqref{eq:SS3} as: $\int_a^b Ap\exp(-Bp/E(r))\dd{r}=\ln(1+1/\gamma)$. The proposed formalism here includes the volume change as the electrons move from the inner to the outer electrode via the additional mobility term: $\ln(\mu(E_a)/\mu(E_b))\approx D\ln((b/a)^\delta)$.

Section~\ref{sec:Model} has established the equivalence between the classic theory for parallel plate electrodes and the equations developed in this work. In addition, the revised equations are fully consistent with the well-established scaling laws. We further note that all the geometrical parameters in Equation~\eqref{eq:SS4} appear in a product with $p$ (i.e., $pa$, $pb$, $pr$, and $pd$). Consequently, one must have $E_{\rm cr}\propto p$ so that Equation~\eqref{eq:SS4} remains true if the pressure changes with all other parameters remaining the same. Similarly, Equation~\eqref{eq:Vcr} establishes the invariance of the critical voltage $V_{\rm cr}$ through pressure changes. Writing $E_{\rm cr}=\dfrac{E_{\rm cr}}{p}p$ lets us rewrite Equations~\eqref{eq:Vcr_Ca}, \eqref{eq:Vcr_Cy}, and \eqref{eq:Vcr_Sp} to display the pressure scaling explicitly as follows:
\begin{linenomath*}
\begin{subequations}\label{eq:Vcr2}
	\begin{align}
	V_{\rm cr}&=\dfrac{E_{\rm cr}}{p}pa\cdot\frac{pd}{pa} & \delta&=0\label{eq:Vcr2_Ca}\\
	V_{\rm cr}&=\dfrac{E_{\rm cr}}{p}pa\cdot\ln\left(1+\dfrac{pd}{pa}\right) & \delta&=1\label{eq:Vcr2_Cy}\\
	V_{\rm cr}&=\dfrac{E_{\rm cr}}{p}pa\cdot\dfrac{\dfrac{pd}{pa}}{1+\dfrac{pd}{pa}} & \delta&=2\label{eq:Vcr2_Sp}
	\end{align}
\end{subequations}
\end{linenomath*}
Since $E_{\rm cr}/p$ is constant, then $V_{\rm cr}$ only depends on the parameters $pd$ (as in the classic \cites{Townsend:1901} theory) and $pa$. The previously established scaling law stands with the additional parameter $pa$. Therefore, Equations \eqref{eq:Ecr} and \eqref{eq:Vcr2} demonstrate that both the critical electric field, $E_{\rm cr}$, and potential, $V_{\rm cr}$, are functions of the reduced electrode and gap sizes, namely $pa$ and $pd$. 

Alternately, the ideal gas law, $p=Nk_{\rm B}T$, allows us to rewrite $E_{\rm cr}$ and $V_{\rm cr}$ as functions of $Nd$ and $Na$, where $N$ is the number gas density. This result holds for constant gas temperature, which is a reasonable assumption for a cold, non-thermalized discharge such as corona or glow. However, it is worth noting that the coefficients $A$, $B$, and $D$ are derived from a fit to the Boltzmann equation using the parameters given in Table~\ref{tab:atm}. The BOLSIG+ solver \citep{Hagelaar:2015} requires a temperature input while it outputs reduced values for $\alpha$ and $\mu$ using the number density $N$. The pressure conversion is necessary to compare to experimental data. The temperatures we used for the four worlds are summarized in Table~\ref{tab:atm}. The conversions of the coefficients from density to pressure call for the neutral gas temperature (see \ref{app:Conversions}) and this information is required for direct comparison between experimental and theoretical Paschen curves. Any modification to the $A$ and $B$ coefficients will primarily shift the curves and surfaces along the vertical $V_{\rm cr}$ and horizontal $pd$ axes of Figures~\ref{fig:Earth} to~\ref{fig:Venus}, respectively. 

In all considered cases, $E_{\rm cr}/p$ and $V_{\rm cr}$ present an asymptotic behavior towards infinite electric field and potential for low values of $pd$, independently of $pa$. The values on the left branches of the Paschen curves ($pd\lesssim pd_{\min{}}$ in column (IV) of Figures~\ref{fig:Earth} through \ref{fig:Venus}) should be taken with caution as they may not describe the physical mechanism occurring in high-electric fields. Discharges at very low $pd$ correspond to dielectric breakdown in a quasi-vacuum. Indeed, \citet[p.~135]{Raizer:1991} noticed that the electron-avalanche process responsible for self-sustained discharges between parallel plates is replaced by cathode emission at $pd\lesssim\unit[10^{-3}]{cm\cdot Torr}$. Thus, the properties of the discharge are no longer defined by the neutral gas between the electrode, but rather by the metallic composition of the electrode. For this reason, the numerical solutions presented in this work do not apply in such regimes. On the other hand, the convergence across gas compositions at high $pd$ indicates that the number density of the neutral gas can become a dominant factor over the molecular electric properties for large gaps under significant pressure. The right branches show similarities to each other across the geometries at large $pd\lesssim\unit[100]{cm\cdot Torr}$. We advance that these similarities (observed in column (IV) of Figures~\ref{fig:Earth} through \ref{fig:Venus}) reflect a regime where the electrode radii of curvature are large enough relative to the gap sizes to result in quasi-plane-to-plane conditions.


Considering the uncertainty of the electrode geometries in experimental data, the theoretical curves are consistent with the measurements. Both approaches indicate minima in the critical voltages around $\unit[0.5]{cm\cdot Torr}$. A simple division by the atmospheric pressure returns the gap size most susceptible to trigger a self-sustained discharge in a given atmosphere. For example, under a pressure of \unit[760]{Torr} (Earth's surface pressure), dielectric breakdown may occur in gaps sizes $\approx\unit[5]{\mu m}$ at $V_{\rm cr}\lesssim\unit[500]{V}$. At tropopause's levels, $p\approx\unit[200]{Torr}$ and the same voltage can initiate a Townsend discharge in a larger gap ($d\approx\unit[25]{\mu m}$). Similarly, the lower pressure in the Martian atmosphere indicates that larger gaps are more prone to dielectric breakdown at the surface of Mars.

Panels (f) and (j), i.e., Column (II), of Figures~\ref{fig:Earth} to~\ref{fig:Venus} emphasize the added role of mobility in non-planar geometries. In particular, these plots suggest that a reduced radius $pa$ of $\approx\unit[1]{cm\cdot Torr}$ is better for initiating self-sustained glow discharges. This corresponds to radii of curvatures $a\approx\unit[0.05/1]{mm}$ for Earth and $\unit[0.2/5]{mm}$ for Mars at ground and cloud levels ($z\approx\unit[10/20]{km}$) in the atmosphere. Such radii of curvature are consistent with previous theories that sharp-tipped rods should facilitate the initiation of upward-connecting leaders and result in better lightning protection, a prediction contrary to field studies \citep[e.g.,][]{Moore:1983,Moore:2000b,Moore:2000a,Moore:2003}. The paradox therefore remains. However, for Earth, the previous calculations indicate that millimeter-sized ice graupels are in the ideal size range for discharge initiation at cloud altitude. Beyond meteorological multiphase flows, numerous Earth systems transport particles in these size ranges and involve discharges processes across a wide range of scales \citep{Crozier:1964, Kamra:1972, Farrell:2004, Cimarelli:2022, mendez2022lifetime}. Such flows include gravity currents (dust storms, pyroclastic density currents), volcanic eruption plumes, and wildfire smoke clouds. Particles in these contexts may have substantial inertia and granular temperatures such that transient optimal gap distances between particles should be common even in very dilute flows \citep{dufek2012granular, dufek2007suspended}. Lastly, if  corona discharge is indeed a precursor to connecting leaders, they can be involved in the process of initiation of lightning and Transient Luminous Events (TLEs), jets and sprites in particular.

While this study provides a framework to constrain the capacity of charged surfaces to cause a breakdown on Mars, Titan, and Venus, how surfaces become electrified on these worlds remains an area of active research. Mars, for instance, lacks a hydrological cycle to drive meteorological electrical activity analogous to that on Earth. While Titan and Venus do have ``hydrological" processes that involve the cycle of  hydrocarbons and sulfur compounds, respectively, whether or not clouds of these compositions are propitious for discharges remains unknown \citep{Hayes:2018, Shao:2020}. 

However, as is the case for Earth, the three other worlds considered here do host granular reservoirs that could provide pathways for non-meteorological discharges \citep{Thomas:1985Dust, Balme:2006, radebaugh2008dunes, Lorenz:2014}. Martian dust storms involve the movement of large amounts of silicate particles which invariably undergo collisions. These interactions could charge dust particles via frictional and contact electrification--collectively known as \textit{triboelectrification} \citep{Horanyi:2001, Melnik:1998, Merrison:2004, Delory:2011}. Indeed, a broad range of experimental efforts suggests that tribocharging may be quite efficient within Martian dust events \citep[e.g.,][]{Eden:1973, Krauss:2003, Wurm:2019, Mendez:2021a}. While these experiments have investigated electrification at the grain scale, computer simulations (sometimes combined with experiments \citep[e.g.,][]{Harrison:2016}) have provided useful full-scale expedients for studies of dust devil electrification \citep[e.g.,][]{Melnik:1998,Farrell:2003}. The results of these studies all converge to the conclusion that  electrification in Martian dust storms should suffice to produce gas breakdown and an atmospheric electric circuit \citep{Farrell:2001}. 

Similar charging processes may operate on Titan and Venus (or any other world with mobile granular materials). On Titan, triboelectrification has been associated with the transport of wind-blown hydrocarbon sand \citep{MendezHarper:2017} and the aggregation of fine photochemical hazes. Very little work has explored triboelectrification under relevant Venusian conditions. However, the presence of dunes and volcanic features on Venus indicates particles may charge frictionally during aeolian transport and eruptions \citep{james2008electrical}.

Determining the conditions under which atmospheric discharges occur has implications for atmospheric chemistry and habitability. Furthermore, discharges could present risks to landers and rovers or cause artifacts that confound the interpretation of sensor data \citep{Krauss:2003}. Recently, for instance, calculations performed by \cite{farrell:2021} suggest that the rotors of the Ingenuity helicopter could cause localized breakdown during landing or takeoff. While videography of initial flights has not revealed visual evidence for discharges, such events may be best detected electronically. Unfortunately, Ingenuity does not have the instrumentation to make such measurements. The upcoming Dragonfly rotorcraft mission to Titan, however, will involve an electric field measurement system (EFIELD) in its DraGMet suite. The main objective of the EFIELD experiments is to measure Schumann resonances, which, if detected, would provide evidence for lightning. Beyond ELF modes, \cite{Chatain:2022} have made a compelling case that the sensor could be used to detect the movement of charged hydrocarbon sand flying past or impinging on the probe during ``brownout" conditions. Because (by definition) discharges also involve the movement of charge, the EFIELD instrument could, in principle, detect small-scale breakdown in he vicinity of the rotorcraft. In the case of Venus,  near-term investigations of discharges in the Venusian environment will remain limited to remote sensing observations and analog experiments. 


\section{Conclusions}\label{sec:Conclusions}
The principal results and contributions from this work can be summarized as follows:
\begin{enumerate}
\item The theoretical treatment of self-sustained discharge between coaxial cylinders or concentric spheres requires a model of mobility. The reduced electron mobility in telluric world atmospheres approximately follows a power law: $\tilde{\mu}\times N = C(E/N)^D$, where $C$ and $D$ are gas-specific constants derived from a numerical fit to the curve $E/N$ vs. $\mu\times N$.
\item The newly proposed formalism explains the slope of the Paschen curves in non-planar geometries and maintains the scaling laws established by the classic theory.
\item In cylindrical and spherical cases, both electrode and gap sizes define the condition of discharge initiation. Thus, Paschen curves and Stoletov's points become surfaces and curves, respectively. 
\item Critical voltages occur at $pd$ and $pa$$\approx$$\unit[0.5]{cm\cdot Torr}$, suggesting easier initiation around millimeter-size particles in dust and water clouds.
\item Glow corona formation is easier in Mars low pressure, $\rm CO_2$-rich atmosphere than in Earth's high-pressure atmosphere.
\end{enumerate}

The specific values of the fit coefficients need revising based on laboratory experiments rather than numerical experiments (i.e., solution of the Boltzmann equation) and will be addressed in future work.

\appendix
\section{Density vs. pressure scaling} \label{app:Conversions}
Experimental work typically adopts pressure-scaled variable, $E/p$, $\alpha/p$, $\mu\times p$ (e.g., columns (III) and (IV) in Figures~\ref{fig:Earth} to~\ref{fig:Venus}), while numerical solvers conventionally prefer the number density $N$ as the scaling factor. Calculations of the fit coefficients $A$, $B$, $C$, and $D$ are performed using numerical solutions but require conversion for comparison with the peer-reviewed experimental data. Equations \eqref{eq:a_p} to \eqref{eq:E_p} provide the conversion factors.

\begin{linenomath*}
\begin{align}
\dfrac{\alpha}{p} &= \left(\dfrac{101325}{100\cdot 760}\cdot\frac{1}{k_{\rm B}T}\right)\dfrac{\alpha}{N} \label{eq:a_p}\\
\mu\times p &= \left(\dfrac{10^4\times760}{101325}\cdot k_{\rm B}T\right)\mu\times N \label{eq:uxN}\\
\frac{E}{p} &= \left(\frac{101325\cdot 10^{-21}}{100\cdot 760}\cdot\dfrac{1}{k_{\rm B}T}\right)\frac{E}{N} \label{eq:E_p}
\end{align}
\end{linenomath*}
where the variables have the units indicated in parentheses: $\unit[\alpha/p]{\left(1/\left(cm\cdot Torr\right)\right)}$, $\unit[\mu\times p]{\left(\left(cm^2\cdot Torr\right)/\left(V\cdot s\right)\right)}$, $\unit[E/p]{\left(V/\left(cm\cdot Torr\right)\right)}$, $\unit[\alpha/N]{\left(m^2\right)}$, $\unit[\mu\times N]{\left(1/\left(V\cdot m\cdot s \right)\right)}$, $\unit[E/N]{\left(Td\right)}$, $\unit[k_{\rm B}]{\left(J/K\right)}$, and $\unit[T]{\left(K\right)}$, respectively.

Similarly, the fit coefficients $A$ and $B$ need converting. If indices $p$ and $N$ indicate the variables used for density and pressure calculations, then:
\begin{linenomath*}
\begin{align}
A_p &= \left(\dfrac{101325}{100\cdot760}\cdot\frac{1}{k_{\rm B}T}\right)A_N \\
B_p &= \left(\dfrac{101325\cdot10^{-21}}{100\cdot760}\cdot\frac{1}{k_{\rm B}T}\right)B_N
\end{align}
\end{linenomath*}
The coefficient $D$ is unchanged, while $C$ cancels out from the equations and requires no conversion. 

\section*{Acknowledgment}
This work was supported by the National Science Foundation (grant number: 2047863), Embry-Riddle Aeronautical University Office of Undergraduate Research.

\bibliographystyle{Files/elsarticle-harv}

\end{document}